\documentclass[a4paper,nobibnotes,nofootinbib,notitlepage,superscriptaddress]{revtex4-1}

\usepackage{amssymb}
\usepackage{amsmath}
\usepackage{epsfig}
\usepackage{graphicx}

\newcommand{\be}{\begin{equation}}
\newcommand{\ee}{\end{equation}}
\newcommand{\bea}{\begin{eqnarray}}
\newcommand{\eea}{\end{eqnarray}}

\newcommand{\pom}{\mathbb{P}}
\newcommand{\reg}{\mathbb{R}}

\begin{document}

\title{Prompt photon production in double-Pomeron-exchange events at the LHC}

\author{A.K. Kohara}\email{kendi@if.ufrj.br}
\affiliation{Centre de Physique Th\'eorique, \'Ecole Polytechnique, CNRS, Universit\'e Paris-Saclay, F-91128 Palaiseau, France}
\affiliation{Instituto de F\'isica - Universidade Federal do Rio de Janeiro, C.P. 68528, Rio de Janeiro 21945-970, RJ, Brazil}

\author{C. Marquet}\email{cyrille.marquet@polytechnique.edu}
\affiliation{Centre de Physique Th\'eorique, \'Ecole Polytechnique, CNRS, Universit\'e Paris-Saclay, F-91128 Palaiseau, France}

\begin{abstract}

Within the resolved Pomeron model of hard diffractive scattering, we compute prompt photon production in double-Pomeron-exchange events in proton-proton collisions. Using specific kinematical constraints chosen according to the acceptances of the forward proton detectors of experiments at the Large Hadron Collider, we provide estimates for inclusive and isolated photon production. This is done using the JetPhox program. We find that next-to-leading order corrections to the hard process are important and must be included in order to correctly constrain the quark and gluon content of the Pomeron from such processes at the LHC.

\end{abstract}

\maketitle

\section{Introduction}

A large effort has been devoted to understand the QCD dynamics of hard diffractive events in hadronic collisions, since such processes were first observed at HERA \cite{Derrick:1993xh,Ahmed:1994nw} and at the Tevatron \cite{Abachi:1994hb,Abe:1994de} more than 20 years ago. Describing diffractive processes in QCD had been challenging for decades, but the presence of a large momentum transfer in these events brought hope that one could be able to understand them with weak-coupling methods. However, while many years of phenomenology point to the existence of a colorless object called the Pomeron, responsible for diffractive events when exchanged in the $t$-channel, understanding hard diffraction in QCD and describing the Pomeron as a structure composed of quarks and gluons remains a challenge.

In the case of deep inelastic scattering (DIS) $\gamma^*p\!\to\! X$, where leptons collide with protons at high energies through the exchange of a high-virtuality photon, the situation has reached a satisfactory level. Due to several years of experimental efforts at HERA, the diffractive part of the deep inelastic cross-section, which corresponds to about 10\% of the events, has been measured with good accuracy \cite{Chekanov:2008fh,Aaron:2012ad,Aaron:2012hua}. On the theoretical side, the collinear factorization of the DIS cross section also holds for its diffractive component \cite{Collins:1997sr}, which allows to separate the short-distance partonic cross section computable in perturbation theory from the long-distance dynamics encoded in diffractive parton distribution functions (pdfs).

By contrast, the description of hard diffraction in hadron-hadron collisions still poses great theoretical problems. Indeed, Tevatron data provided evidence that even at very large momentum scales, collinear factorization does not apply in such cases \cite{Affolder:2000vb}. In order to estimate hard diffractive cross sections when factorization does not hold, (a modern version of) the resolved Pomeron model \cite{Ingelman:1984ns} is being widely used. It makes use of the diffractive pdfs extracted from HERA, which give the distribution of quarks and gluons inside the Pomeron depending on the $x$ and $Q^2$ kinematical variables, while modeling the additional soft interactions that violate factorization. To better test the validity of this model, and to better understand the Pomeron structure, it is essential to find sensitive observables to be measured in the current colliders experiments.

One way to constrain quarks and gluons inside the Pomeron is to measure prompt photons in diffractive proton-proton (p+p) collisions, as was suggested in \cite{Marquet:2013rja}. However, this study relied on leading-order matrix-elements, since the Forward Physics Monte Carlo generator \cite{fpmc} was used. Subsequent works also relied on LO matrix elements \cite{Mariotto:2013kca}. In this letter, we want to investigate the effects of higher-order corrections, and their impact for a center-of-mass energy of 13 TeV at the LHC, and we shall use instead the JetPhox Monte Carlo \cite{jetphox} to compute the matrix elements at leading order (LO) and at next-to-leading order (NLO).

On the theoretical side, prompt photons refer to high-$p_t$ photons created in a hard process, either directly (direct photons) or though the fragmentation of a hard parton (fragmentation photons) \cite{Owens:1986mp}. On the experimental side, inclusive and isolated photons denote prompt photons measured without or with an isolation cut, respectively. These are two observables that we shall estimate for double-Pomeron-exchange (DPE) events - meaning diffractive p+p collisions from which both protons escape intact - taking into account the kinematical constraints of the forward proton detectors of the CMS-TOTEM Collaborations, or those to be installed by the ATLAS Collaboration in the future \cite{afp}.
 
The plan of the letter is as follows. In section 2, we recall the resolved Pomeron model and its ingredients, explain how to obtain diffractive pdfs that take into account the restricted phase space of the outgoing protons, and describe how these effective diffractive pdfs are used together with the JetPhox program in order to compute prompt photon production in DPE processes in p+p collisions. In section 3, we present our results for a center-of-mass energy of 13 TeV at the LHC, while analyzing the contributions of Compton, annihilation and fragmentation processes, at LO and NLO. Section 4 is devoted to conclusions.

\section{Theoretical formulation}

\subsection{Resolved Pomeron model}

\begin{figure}
\includegraphics[scale=0.22]{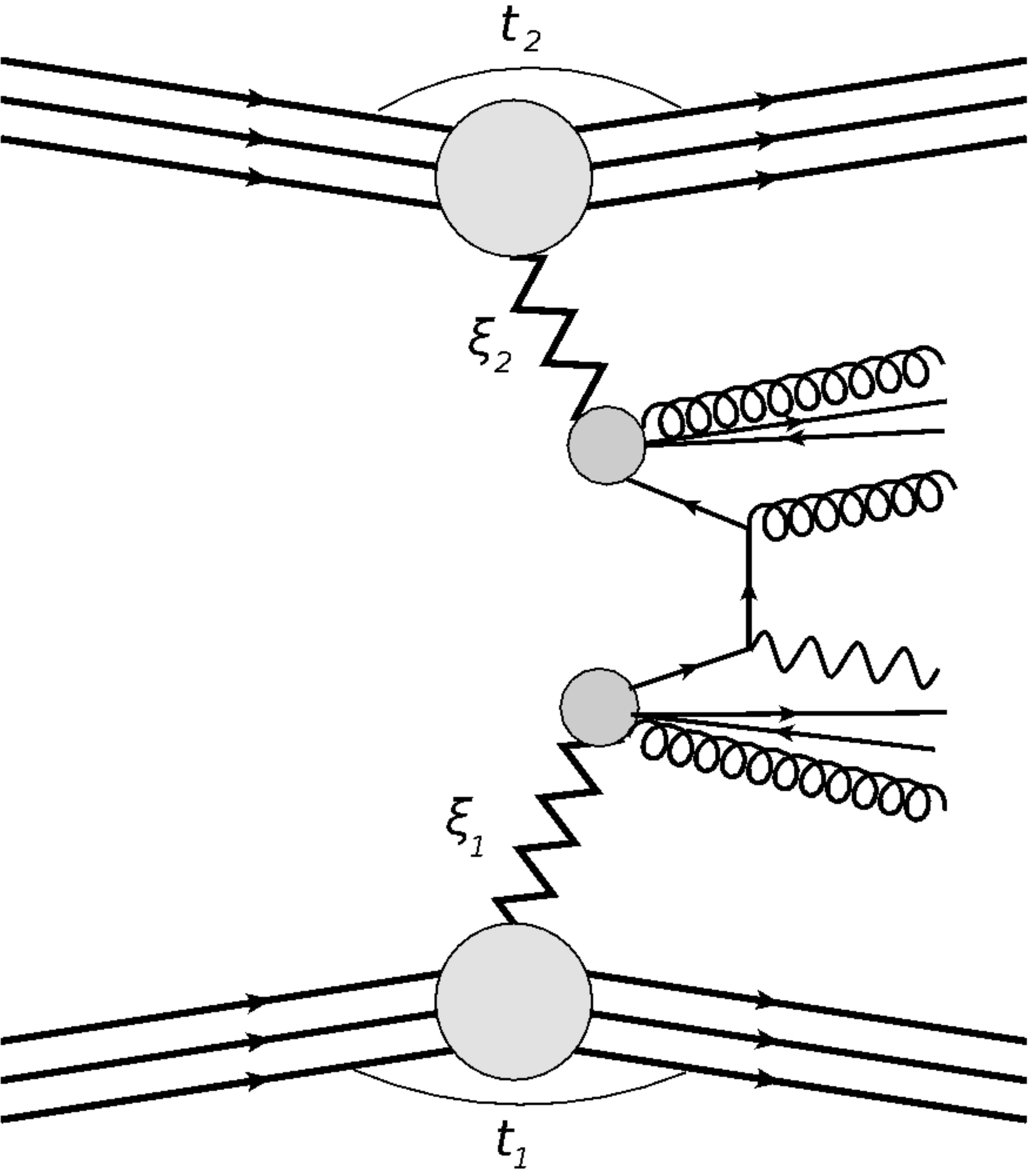}
\hspace{2cm}
\includegraphics[scale=0.22]{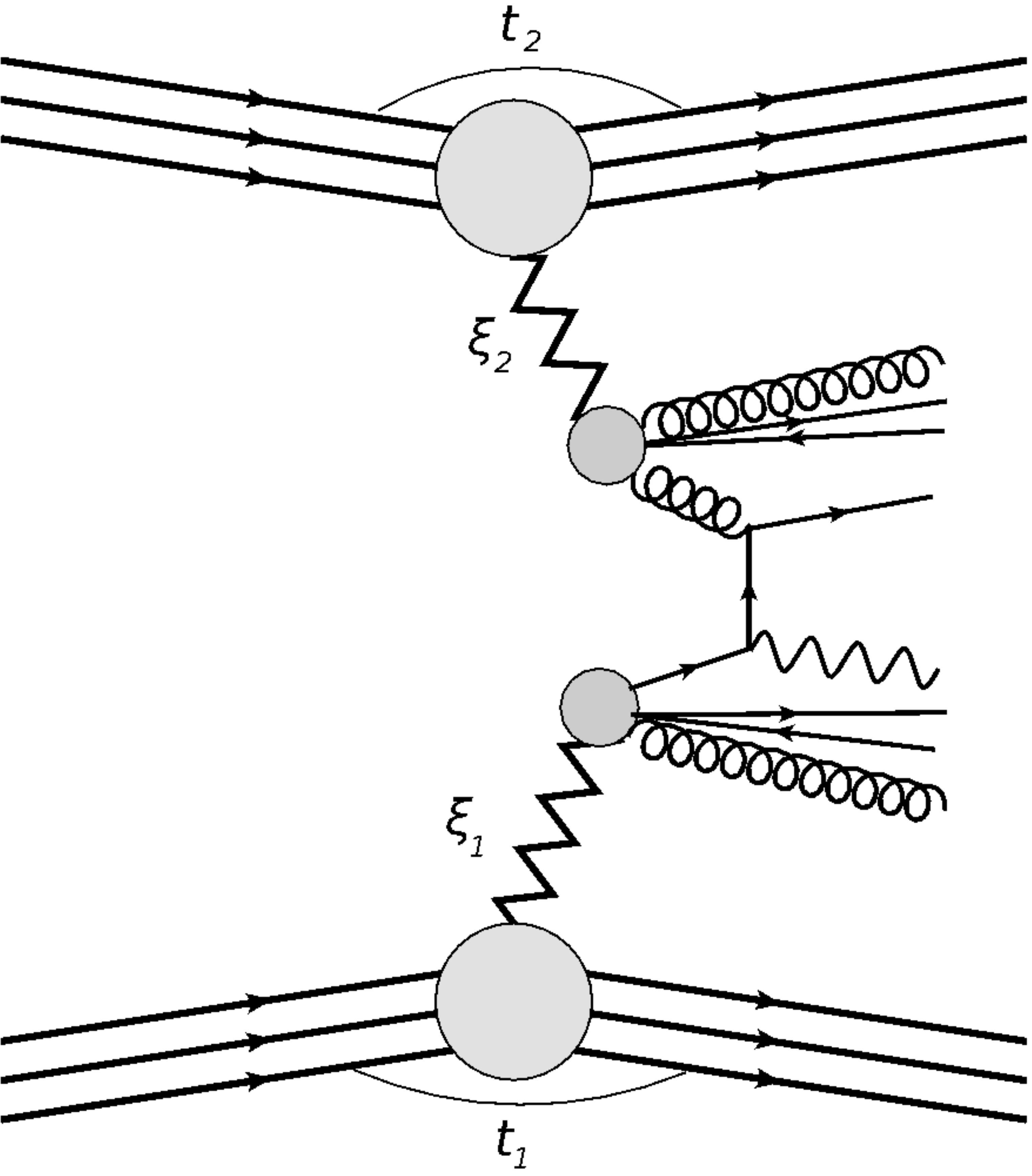}
\caption{Leading-order diagrams for prompt photon production in double-Pomeron-exchange events in p+p collisions. Left: the annihilation partonic sub-processes are only sensitive to the quark content of the Pomeron. Right: the Compton partonic sub-processes are sensitive to the gluon content of the Pomeron as well.}
\label{dpe-photons}
\end{figure}

The resolved Pomeron model is a long-distance/short-distance collinear factorization framework commonly used to calculate hard single-diffraction and DPE processes. In this work, we focus on DPE prompt photon production at the LHC. Related processes (diffractive production of virtual photons and Z bosons) have also been considered previously \cite{Cisek:2009hp,Kubasiak:2011xs}. In the case of direct photons, the leading-order diagrams for this process are pictured in Fig.~\ref{dpe-photons}, and the cross section in the resolved Pomeron model reads:
\be
d\sigma^{pp\to p\gamma Xp}={\cal S}_{DPE}\ \sum_{i,j} \int f^D_{i/p}(\xi_1,t_1,\beta_1,\mu^2)f^D_{j/p}(\xi_2,t_2,\beta_2,\mu^2)\
\otimes d\hat\sigma^{ij\to \gamma X} \label{colfact}
\ee
where $d\hat\sigma$ is the short-distance partonic cross-section, which can be computed
order by order in perturbation theory (provided the transverse momentum of the photon is
sufficiently large), and each factor $f^D_{i/p}$ denotes the diffractive parton distribution in a proton.
These are non-perturbative objects, however their evolution with the factorization scale $\mu$ is obtained perturbatively using the
Dokshitzer-Gribov-Lipatov-Altarelli-Parisi~\cite{dglap} evolution equations.

In \eqref{colfact}, the variables $\xi_{1,2}$ and $t_{1,2}$ denote, for each intact proton, their fractional energy loss and
the momentum squared transferred into the collision, respectively. The convolution is done over the variables $\beta_{1,2}$,
$x_1\!=\!\xi_1\beta_1$ and $x_2\!=\!\xi_2\beta_2$ being the longitudinal momentum fractions of the partons $i$ and $j$ respectively, with respect to the incoming protons. However, hard diffractive cross sections in hadronic collisions do not obey such collinear factorization. This is due to possible secondary soft interactions between the colliding hadrons which can fill the rapidity gap(s). Formula \eqref{colfact} is reminiscent of such a factorization, but it is corrected with the so-called gap survival probability ${\cal S}_{DPE}$ which is supposed to account for the effects of the soft interactions. Since those happen on much longer time scales compared to the hard process, they are modeled by an overall factor, function of the collision energy only. This is part of the assumptions that need to be further tested at the LHC.

In our computations, we shall use diffractive pdfs extracted from HERA data \cite{Aktas:2006hy} on diffractive DIS
(a process for which collinear factorization does hold) by means of an NLO-QCD fit. These are decomposed further into Pomeron and Reggeon fluxes $f_{\pom,\reg/p}$ and parton distributions $f_{i/\pom,\reg}$:
\be
f^D_{i/p}(\xi,t,\beta,\mu^2)= f_{\pom/p}(\xi,t)
f_{i/\pom}(\beta,\mu^2)+f_{\reg/p}(\xi,t) f_{i/\reg}(\beta,\mu^2)\ .
\label{dpdfs}
\ee
The secondary Reggeon contribution is important only at large values of $\xi$, at the edge of the forward proton detector acceptance, and therefore we do not take it into account in the following. Measurements at the LHC will allow to test the validity of this further factorization of the diffractive proton pdfs into a Pomeron flux and Pomeron pdfs, as well as the universality of those Pomeron fluxes and parton distributions.
 
\subsection{Effective diffractive pdfs with experimental constraints} 

In the following, we assume the intact protons in DPE events to be tagged in the forward proton detectors of the CMS-TOTEM Collaborations, or those to be installed by the ATLAS Collaboration in the future \cite{afp}, called AFP detectors. The idea is to measure scattered protons at very small angles at the interaction point and to use the LHC magnets as a spectrometer to detect and measure them. We use the following acceptances \cite{Trzebinski:2014vha}:
\begin{itemize}
\item $0.015 < \xi < 0.15$ for ATLAS-AFP 
\item $0.0001< \xi < 0.17$ for TOTEM-CMS\ .
\end{itemize}

Let us now explain how the diffractive pdfs (\ref{dpdfs}) are constrained by those detector acceptances. We denote the diffractive quark and gluon distributions integrated over $t$ and $\xi$ by $q^D(x,\mu^2)$ and $g^D(x,\mu^2)$ respectively. These effective pdfs are obtained from the Pomeron pdfs  $q_{\mathbb {P}}(\beta,\mu^2)$ and $g_{\mathbb {P}}(\beta,\mu^2)$, and from the Pomeron flux $f_{\mathbb {P}/p}(\xi,t)$.
Let us first integrate the latter over the $t$ variable:
\begin{equation}
\label{flux}
f_{\mathbb {P}}(\xi)=\int_{t_{\rm{min}}}^{t_{\rm{max}}}dt\ f_{\mathbb {P}/p}(\xi,t)\quad\mbox{with}\quad f_{\mathbb {P}/p}(\xi,t)=A_{\mathbb P}\frac{e^{B_\mathbb {P}t}}{\xi^{2\alpha_{\mathbb {P}(t)}-1}}\ . 
\end{equation}
The parameters of Eq.(\ref{flux}) are the slope of the Pomeron flux $B_{\mathbb {P}}=5.5^{-2.0}_{+0.7}$ GeV$^{-2}$, and Pomeron Regge trajectory
$\alpha_{\mathbb P}(t)=\alpha_{\mathbb P}(0)+\alpha_{\mathbb P}'~t$ with $\alpha_{\mathbb P}(0)=1.111 \pm 0.007$ and $\alpha_{\mathbb P}'=0.06^{+0.19}_{-0.06}$ GeV$^{-2}$. The boundaries of the $t$ integration are $t_{\rm{max}}=-m_p^2\xi^2/(1\!-\!\xi)$ ($m_p$ denotes the proton mass) and $t_{\rm{min}}=-1$ GeV$^2$. The normalization factor $A_{\mathbb P}$ is chosen such that $\xi\times\int_{t_{\rm{min}}}^{t_{\rm{max}}}dt~f_{\mathbb {P}/p}(\xi,t)=1$ at $\xi=0.003$.

\begin{figure}
\includegraphics[scale=0.37]{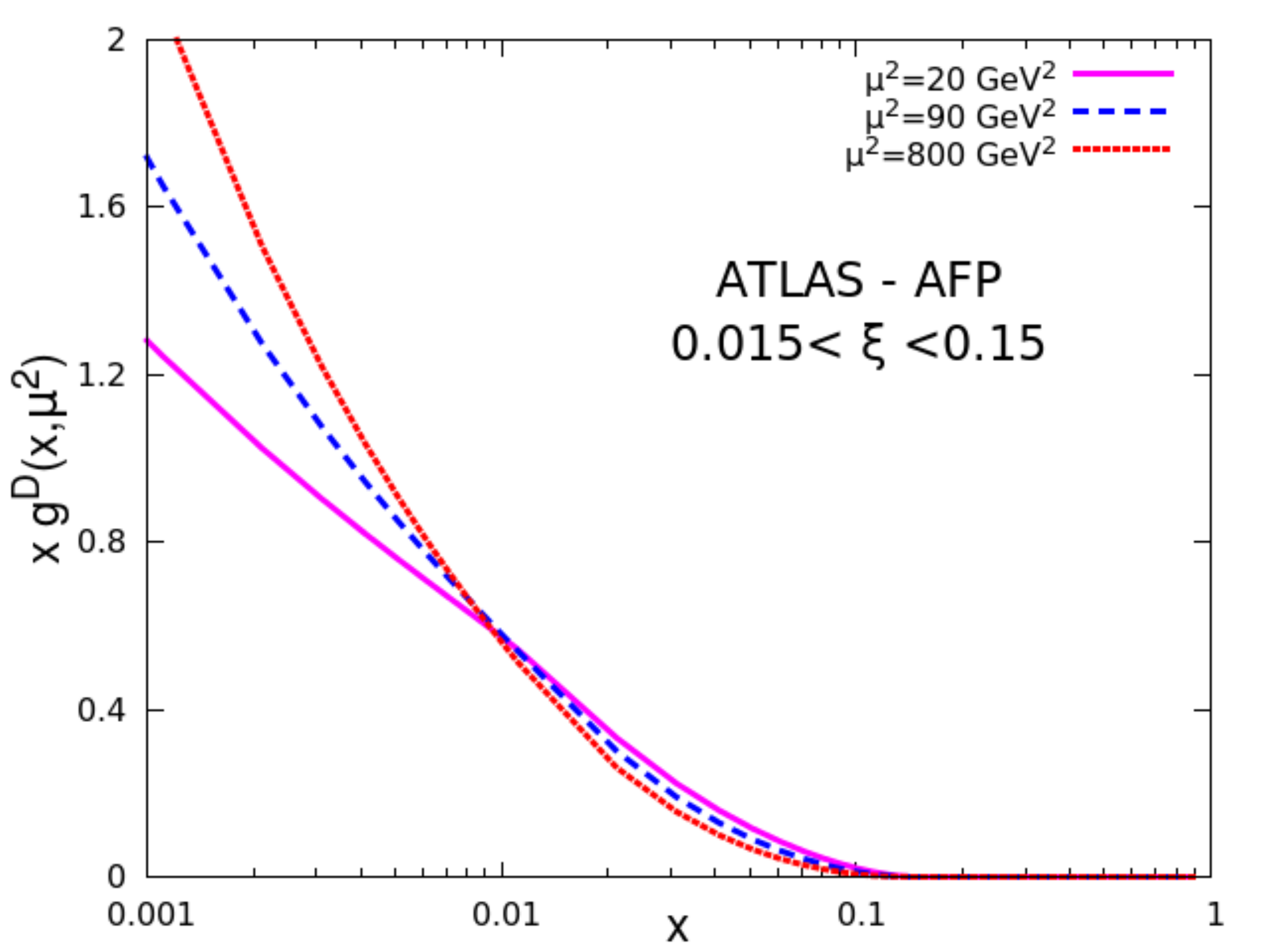}
\includegraphics[scale=0.37]{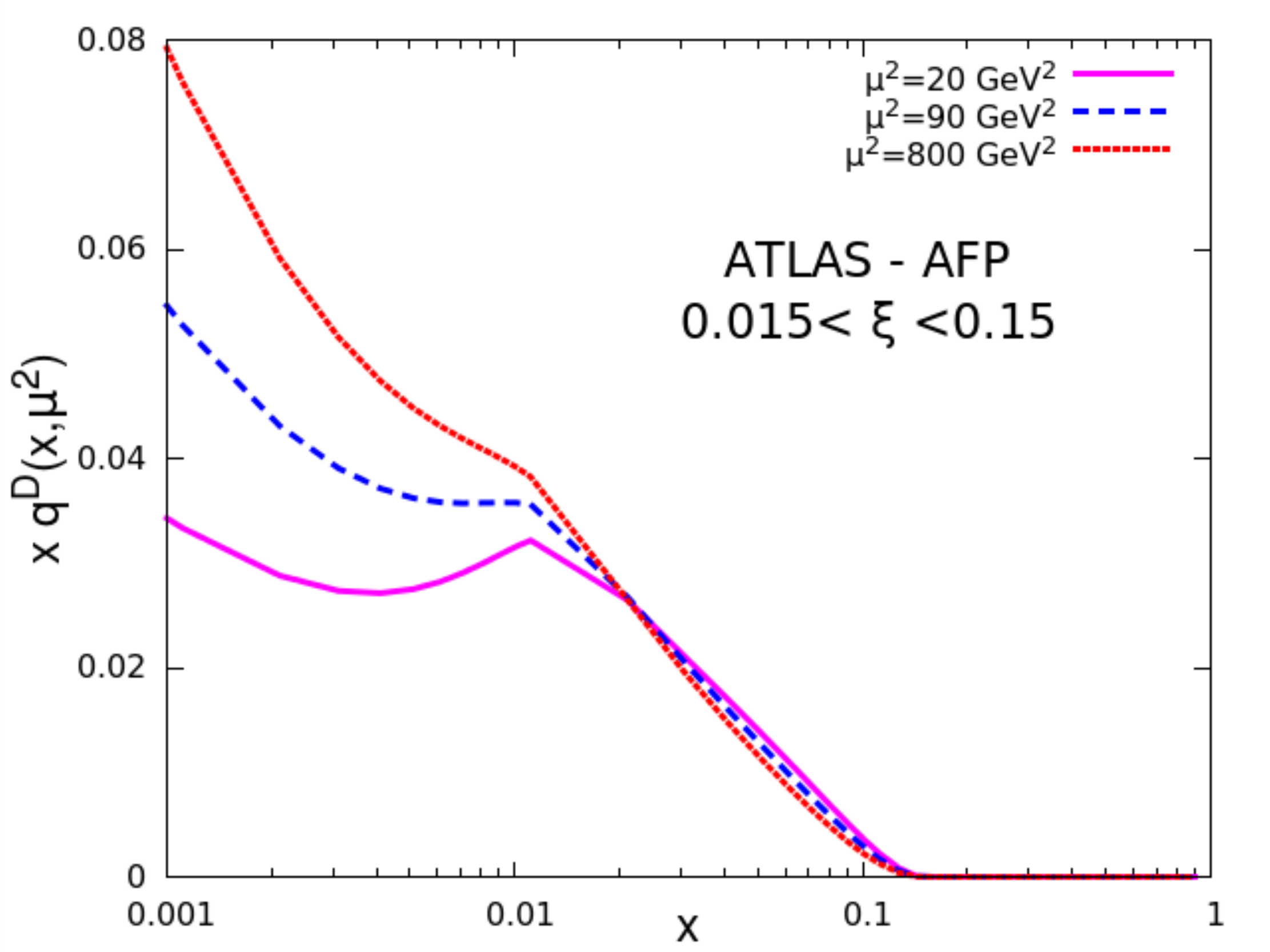}
\includegraphics[scale=0.37]{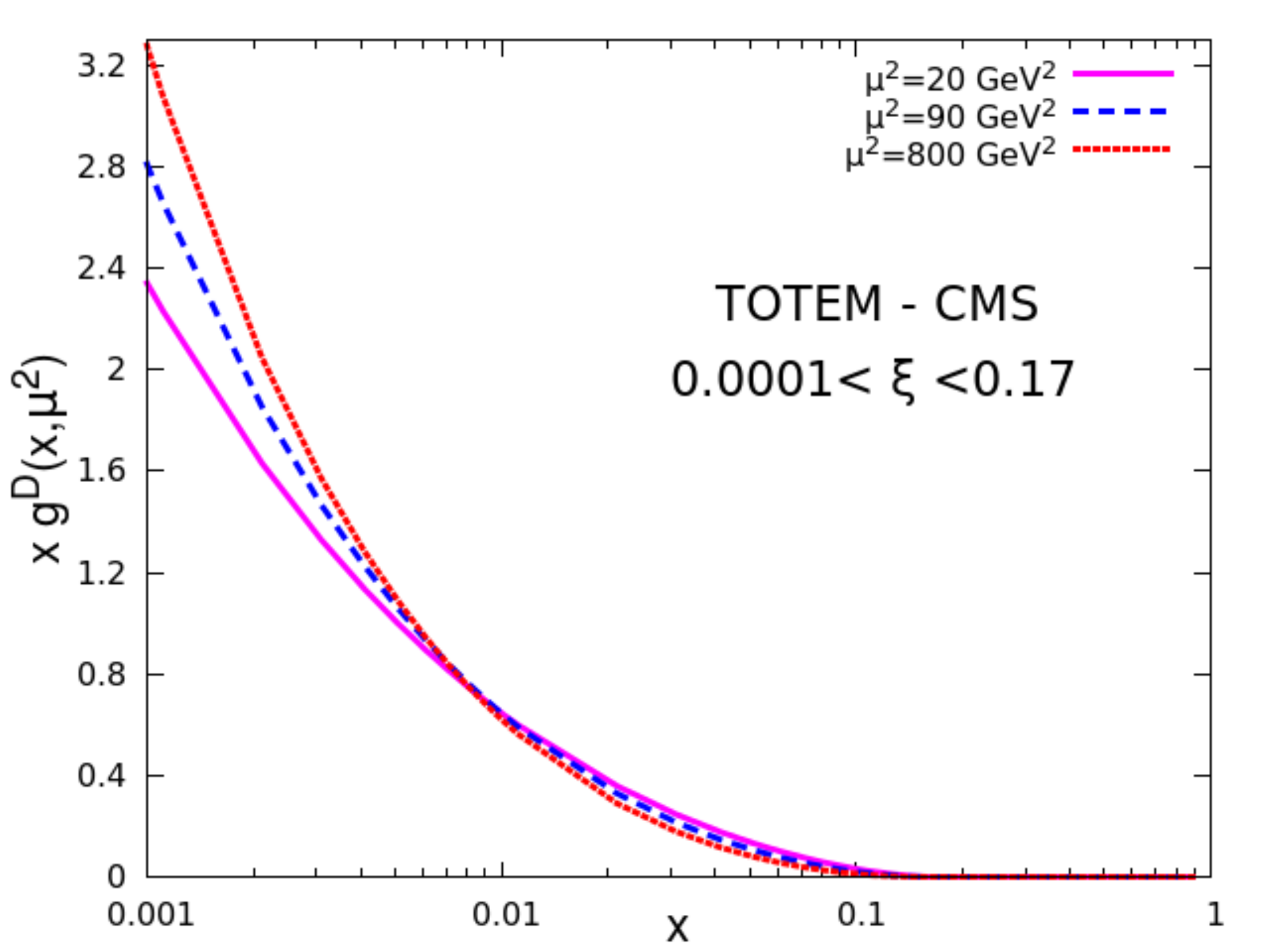}
\includegraphics[scale=0.37]{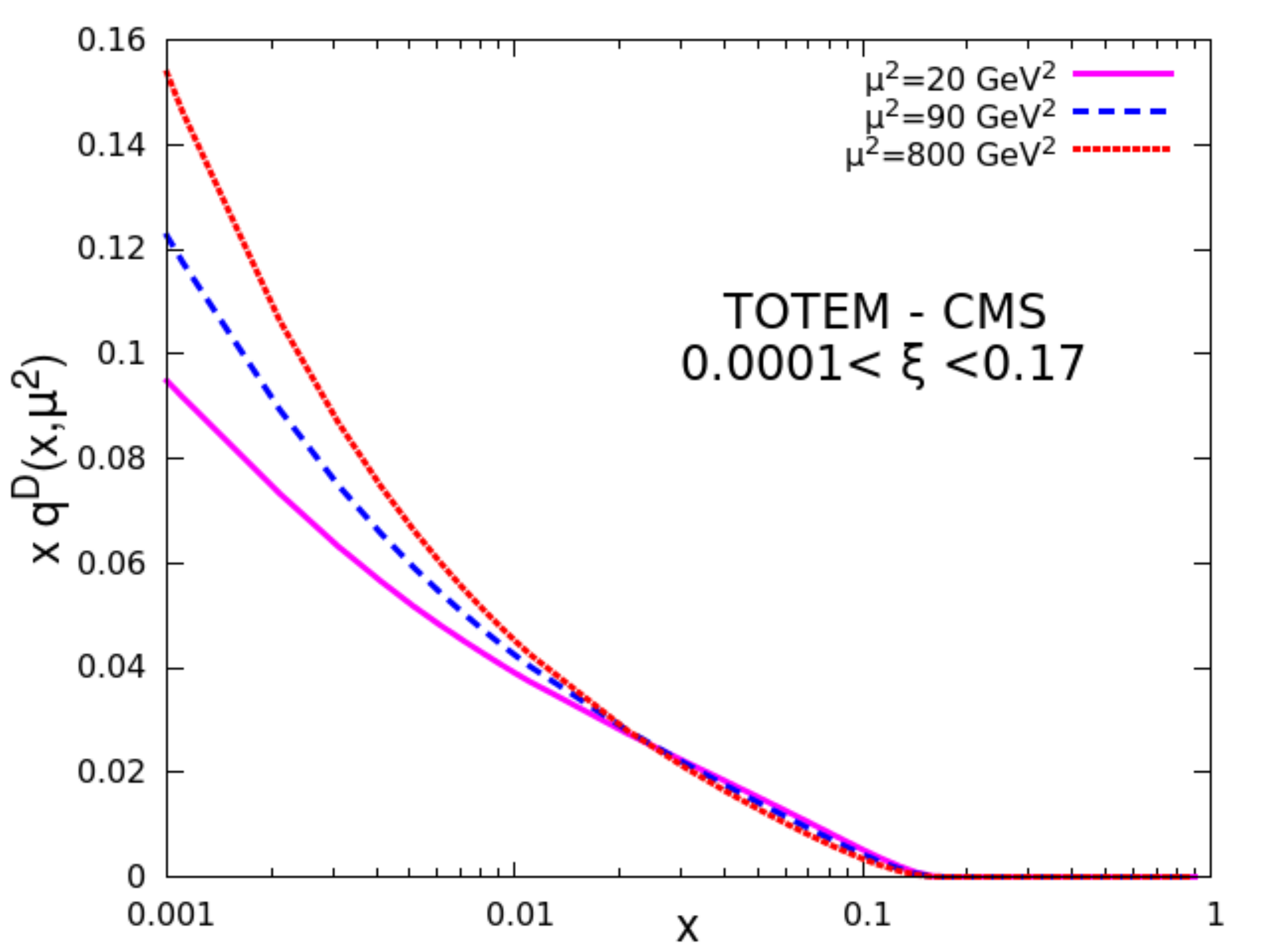}
\caption{These plots represent diffractive proton pdfs for three values of $\mu^2$, with the constraint that the intact proton fall into a forward detector. As a result the distributions vanish for $z>\xi_{\rm{max}}$, and feature a kink when $z$ crosses $\xi_{\rm{min}}$. Left: diffractive gluon distribution $g^D\equiv f^D_{g/p}$ with ATLAS-AFP (top) and TOTEM-CMS (bottom) constraints. Right: the sum of the valence quarks distribution $q^D\equiv \sum_{val} f^D_{q/p}$ with ATLAS-AFP (top) and CMS-TOTEM (bottom) constraints.}
\label{proton_pdfs}
\end{figure}

Next, to obtain the constrained diffractive pdfs, we convolute the Pomeron flux with the Pomeron pdfs while imposing a reduction in the phase space of $\xi$, according to the experimental acceptance of the forward detectors:
\begin{equation}
f^D_{i/p}(x,\mu^2)= \int_{\rm{max}(x,~\xi_{\rm{min}})}^{\rm{max}(x,~\xi_{\rm{max}})}\frac{d\xi}{\xi}~
f_{\mathbb {P}}(\xi)\ f_{i/\mathbb{P}}(x/\xi,\mu^2)\ .
\label{edpdfs}
\end{equation}
For the Pomeron pdfs, we make use of the HERA fit B in \cite{Aktas:2006hy}. In Fig.~\ref{proton_pdfs}, we show the resulting effective diffractive pdfs for both the ATLAS-AFP and TOTEM-CMS constraints. These distributions are built in a way to be easily incorporated into the LHAPDF library \cite{lhapdf} in the grid format.  

\subsection{Computing double-Pomeron-exchange prompt photon production using JetPhox}

JetPhox is a Monte Carlo generator built to compute hadronic cross sections for the process $pp\!\to\!\gamma X$ using the collinear factorization framework. Cross sections are calculated as a convolution of short-range matrix elements, computed at LO and NLO, and long-range (non-perturbative) parton distribution and fragmentation functions. Therefore, within the resolved Pomeron model (\ref{colfact}), this program can also be used to compute the cross-sections $pp\!\to\!pp\gamma X$ in DPE events. In order to do this, we must substitute the regular pdfs by our effective diffractive pdfs:
\begin{equation}
f_{i/p}(x,\mu^2)\longrightarrow \int d\xi dt d\beta\ \delta(x-\beta\xi)\ f^D_{i/p}(\xi,t,\beta,\mu^2)\equiv f^D_{i/p}(x,\mu^2)\ ,
\end{equation}
and multiply the resulting cross sections by the gap survival probability $S_{DPE}$.

JetPhox produces both inclusive and isolated photons with momentum $p_t$ and rapidity $y$. In case of the inclusive cross section, it sums the direct and the fragmentation contribution in the following way:
\be
\frac{d\sigma}{dp_t^2 dy}=\frac{d\hat{\sigma}^{\gamma}}{dp_t^2 dy}+
\sum_{a}\int dz\ \frac{d\hat{\sigma}^{a}}{dp_{ta}^2 dy_a}(p_t/z,y)D_{a}^{\gamma}(z,\mu^2),
\label{diff}
\ee
where $d\hat{\sigma}^{a}$ is the hard cross section for producing a parton $a$=(q,$\bar {\rm q} $,g) which will then radiate a high-$p_t$ photon during its fragmentation into a hadron. $D_{a}^{\gamma}$ is the fragmentation function, the z variable is $z\equiv p_{\gamma}/p_a$, and we have chosen the fragmentation scale to be $\mu$. In case of isolated photons, an additional criteria is imposed on the hadronic activity surrounding the high-$p_t$ photon, as is discussed later.

In the following, we use JetPhox to compute the direct and the fragmentation contributions in (\ref{diff}), replacing, as explained previously, the regular pdfs by the diffractive pdfs extracted above. Technically, this program calls the pdfs from the LHAPDF library \cite{lhapdf}; we replaced one of those parton distribution sets in grid format by our diffractive pdfs constrained with the kinematical cuts. Our choice of factorization scale is $\mu=p_t$.

\section{Numerical results}

In this section, we detail the future measurements to be performed at the LHC, in order to test the resolved Pomeron model and to constrain the quark and gluon content of the Pomeron, using photon production in DPE processes. We use the Monte Carlo program JetPhox (version 1.3.1) to simulate the results, with $2\times10^8$ events per channel.

\subsection{DPE inclusive photons}

\begin{figure}
\includegraphics[scale=0.35]{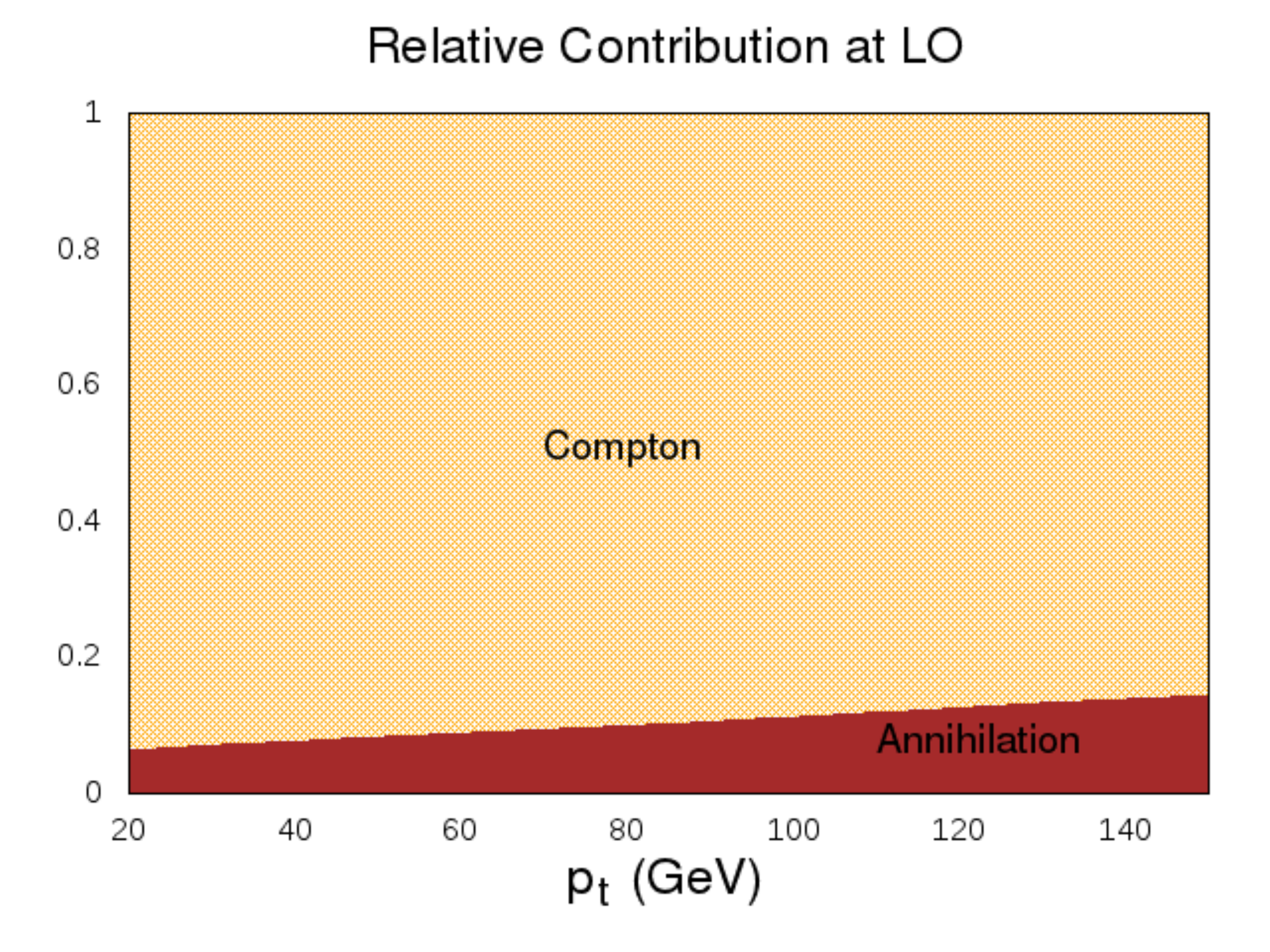}
\includegraphics[scale=0.35]{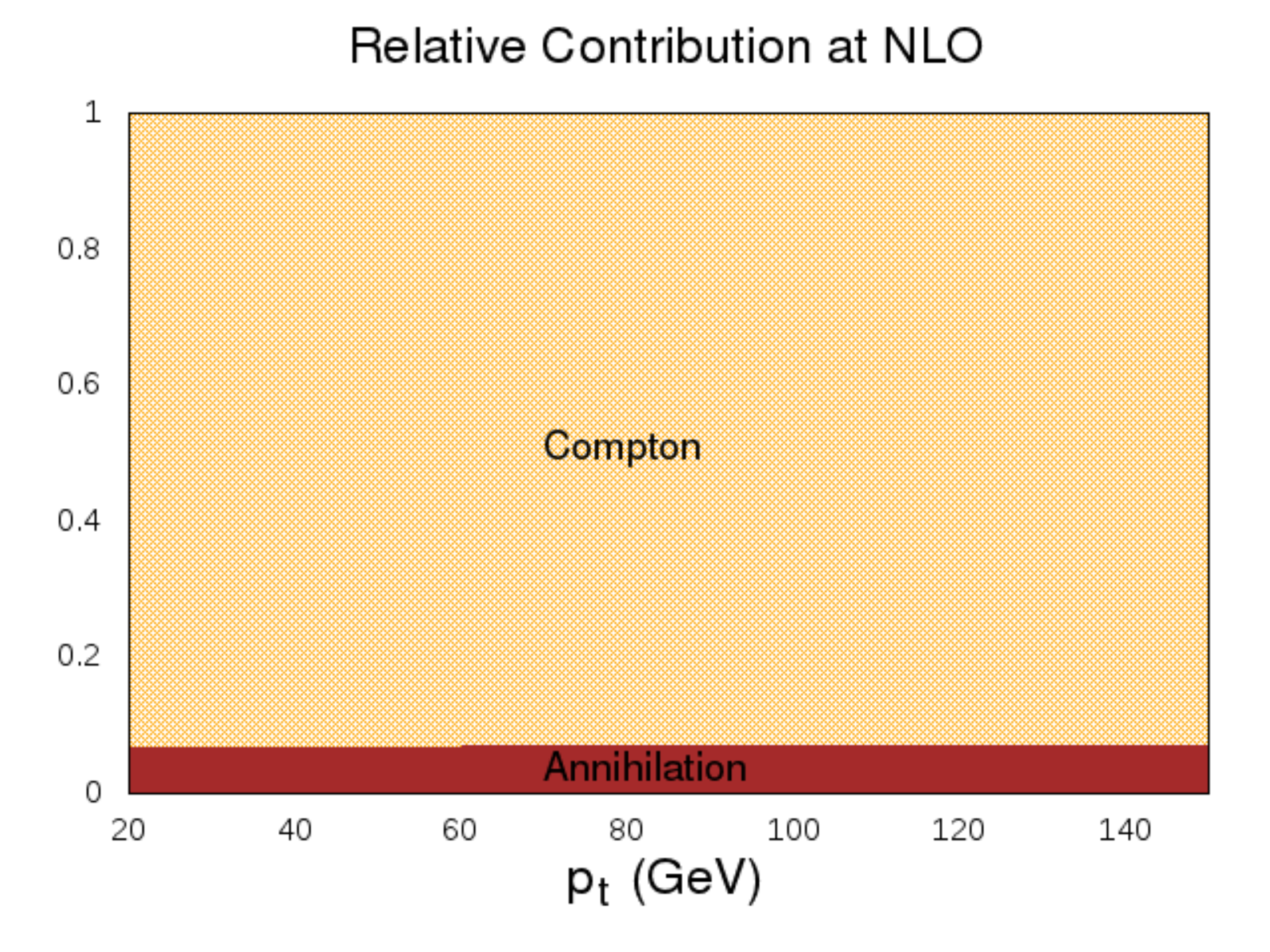}
\caption{These figures show, for DPE direct photon production, the relative contributions of the Compton and annihilation processes at a function of photon $p_t$. Left: at LO, the Compton process represents about 90$\%$ of the differential cross section; the contribution of the annihilation process is slightly increasing with increasing $p_t$. Right: at NLO, the Compton process dominates around 95$\%$ of the differential cross section for all the $p_t$ range analyzed.}
\label{relative_inclusive}
\end{figure}

In the inclusive mode, there are significant contributions from both direct and fragmentation photons; let us first focus on the direct photons.
At LO ($\alpha_{em} \alpha_s$), both annihilation processes $q\bar{q}\to g\gamma$, and Compton processes $q(\bar{q})g\to q(\bar{q})\gamma$, contribute. Going to higher orders opens up additional gluon-initiated sub-processes: $gg\to q\bar{q}\gamma$ (NLO) and $gg\to g\gamma$ (NNLO) (but the latter contributes only to 1$\%$ of the events). Analyzing the relative contributions between the annihilation and Compton channels in DPE events represents a direct way to assess their sensitivity to the quark and/or gluon structure of the Pomeron. This is done in Fig.~\ref{relative_inclusive}, as a function of the photon transverse momentum and using ATLAS-AFP acceptance (very similar results are obtained in the TOTEM-CMS case). We observe a large dominance of the Compton processes, which could be expected considering the relative magnitude of the diffractive gluon and quark distribution shown in Fig.~\ref{proton_pdfs}.

\begin{figure}
\includegraphics[scale=0.4]
{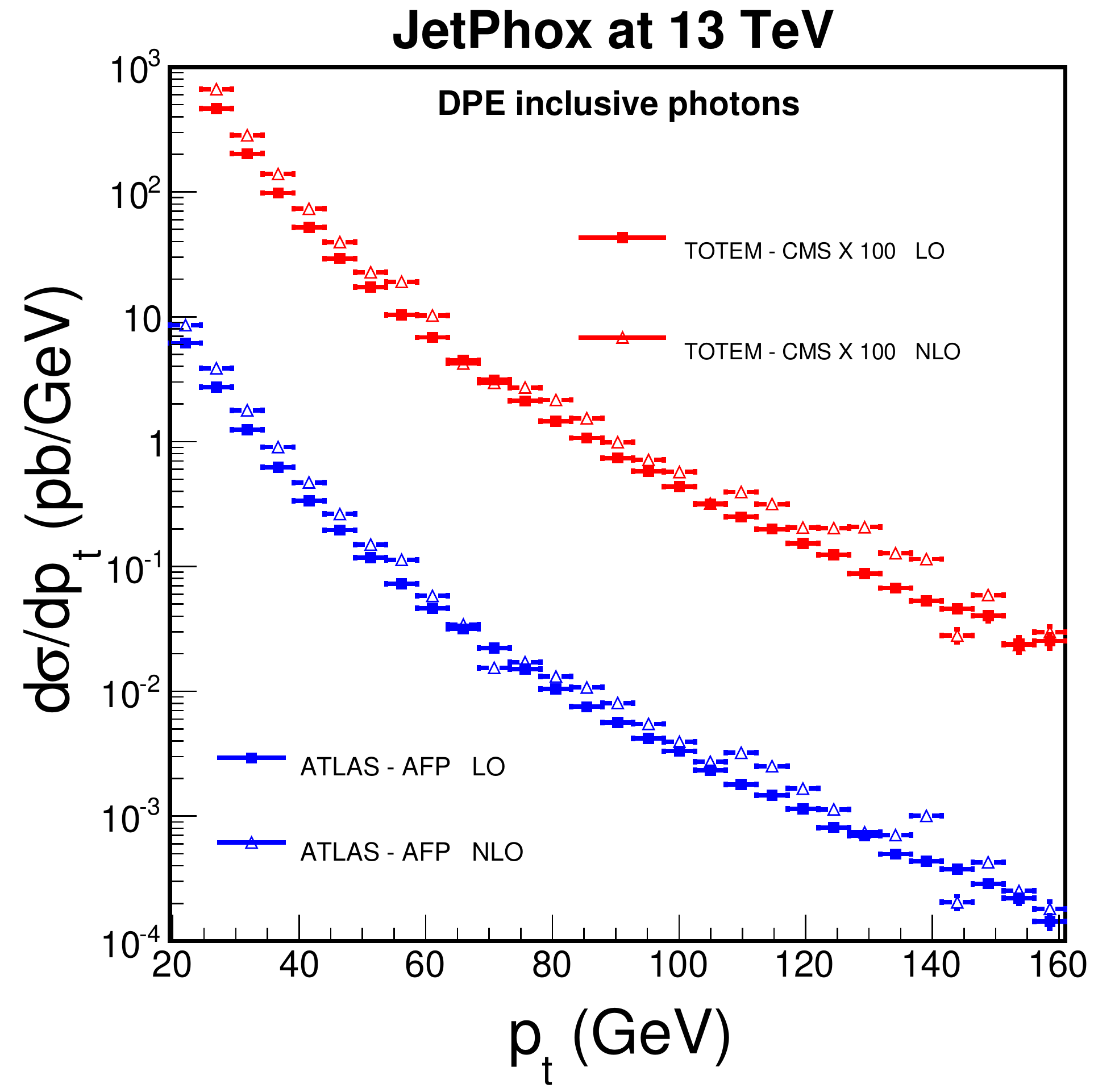}
\includegraphics[scale=0.4]
{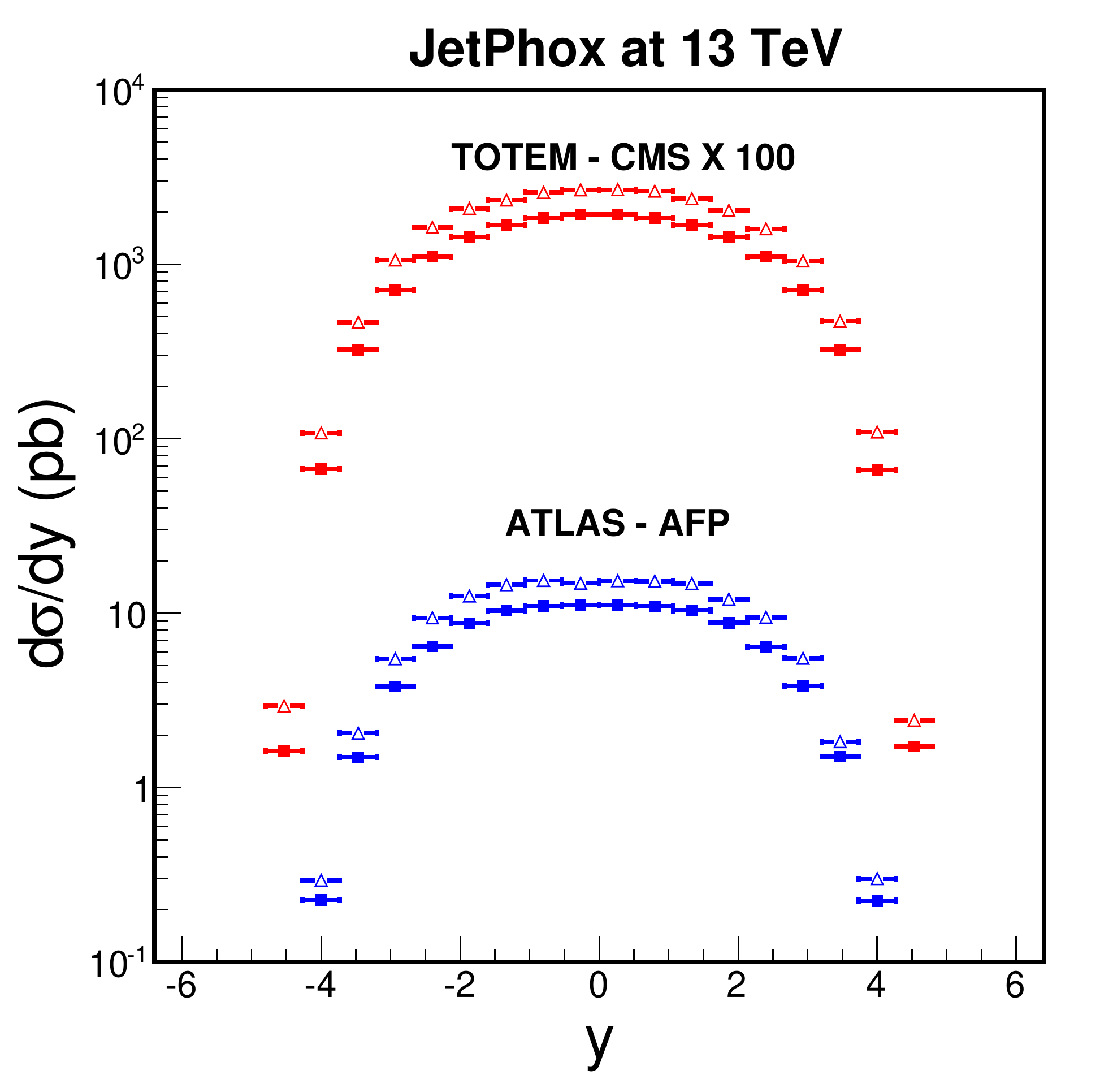}
\caption{These figures show the $p_t$ spectrum (left) and the rapidity distribution (right) of DPE inclusive photons $(pp\to pp\gamma X)$ computed by summing the direct and fragmentation contributions for a center of mass energy of 13 TeV, for both ATLAS-AFP and TOTEM-CMS detector acceptances. The squares and the triangles represent respectively the LO  and NLO calculations, the latter giving cross section about 20$\%$ greater than the former.}
\label{distributions_inclusive}
\end{figure}

This means that extracting Pomeron quark distributions from DPE inclusive photon measurements will first require that the Pomeron gluon content is well constrained (for instance using DPE dijet measurements \cite{Marquet:2013rja}). This is even more so, since fragmentation photons (which contribute to almost half of the inclusive cross section as we will see below) also come mostly from gluon-initiated process. We display in Fig.~\ref{distributions_inclusive} (left) the differential cross section for the production of DPE inclusive photons as a function of the photon $p_t$, summing all the channels and comparing the results at LO and NLO. We show predictions for both ATLAS-AFP and TOTEM-CMS detectors at 13 TeV. In Fig.~\ref{distributions_inclusive} (right), we show the photon rapidity distribution (for $p_t\!>\!20$ GeV), and we note that the difference in magnitude between the LO and NLO calculation is about 20$\%$. Obviously, NLO corrections are not negligible, they must be taken into account in order to extract correctly the Pomeron structure from future data. Note that due to the vanishing of the effective diffractive pdfs (\ref{edpdfs}) for $x\!>\!\xi_{\rm{max}}$, there are no photons produced at very forward or very backward rapidities.

Finally, to compare these cross sections with the future data from the experiments, we note that the gap survival probability factor may have to be readjusted. As advocated in several works \cite{Khoze:2000cy,Kaidalov:2001iz,Bartels:2006ea,Luna:2006qp,Frankfurt:2006jp,Gotsman:2007ac,Achilli:2007pn,Khoze:2008cx}, we have assumed $S_{DPE}\!\simeq\!0.1$, but the actual value is rather uncertain and must first be measured. The general strategy will be the following. First one must perform a global fit of hard diffractive measurements using HERA diffractive DIS data and future LHC DPE dijet data, which, if successful, will constraint well the Pomeron gluon content and determine the actual value of $S_{DPE}$. Then, DPE inclusive and isolated (see below) photon data could also be included in the global fit, which would constrain the Pomeron quark/antiquark content, given the increased sensitivity of such processes to those distributions. Of course, the level of uncertainty will ultimately depend on how precisely such measurements can be performed.

\subsection{DPE isolated photons}

Using the inclusive photon measurement discussed above in order to constrain the quark content of the Pomeron is not optimal, because this observable is contaminated by fragmentation photons, which mostly come from gluon-initiated process. In order to suppress the contribution from fragmentation processes, one can use an isolation criteria that will disregard the photons that are surrounded by too much hadronic activity. Indeed, generically a direct photon will be isolated from a large hadronic activity while a fragmentation photon won't be, since those are collinear emissions from their parent parton. The isolation criteria we use in the following is to require that the hadrons measured within a cone of radius $R=0.4$ have a maximum transverse energy of 4 GeV. This is one of the options available in JetPhox \cite{jetphox}, we checked that our conclusions are independent of this particular choice for the isolation criteria.

\begin{figure}
\includegraphics[scale=0.52]{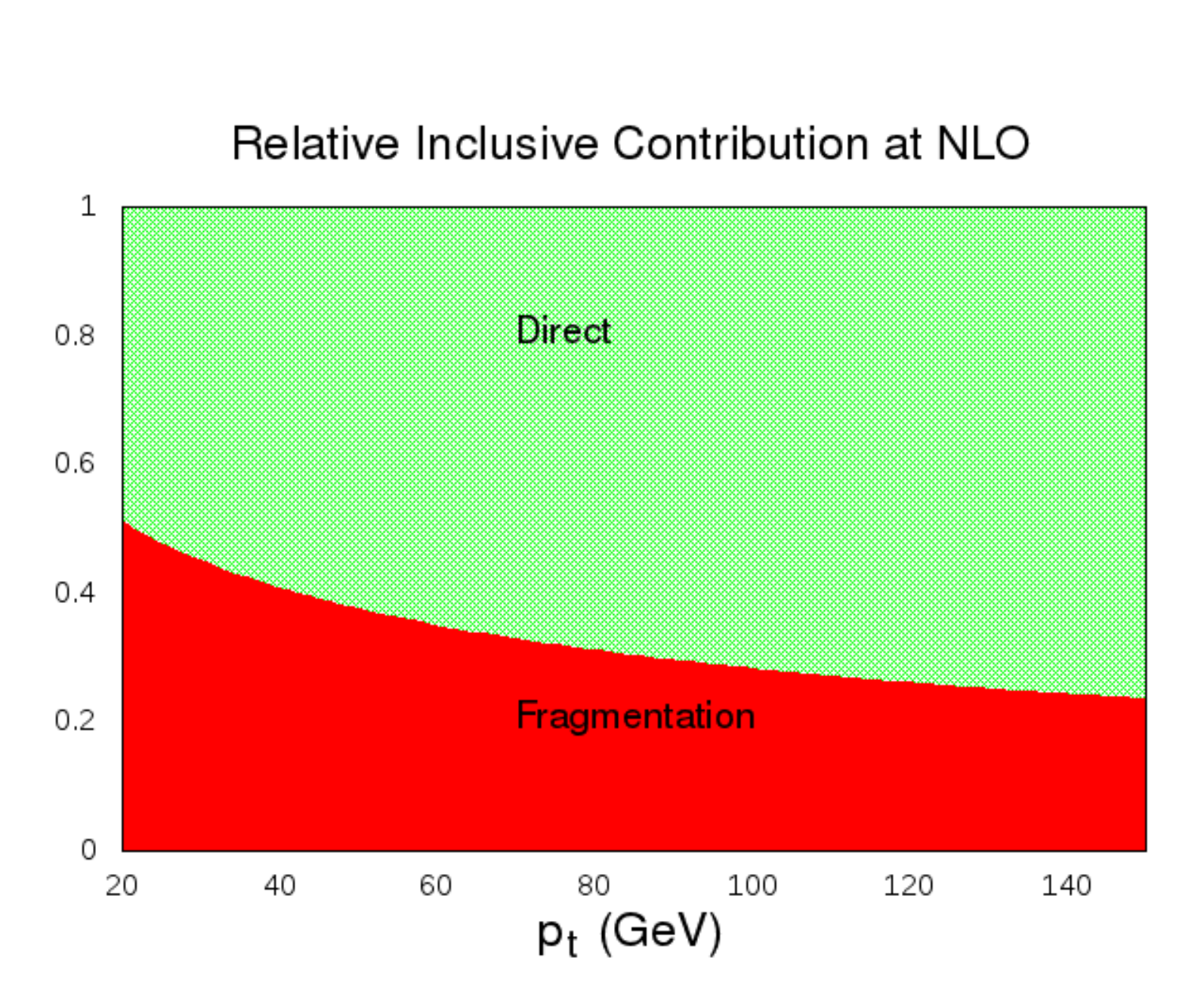}
\includegraphics[scale=0.52]{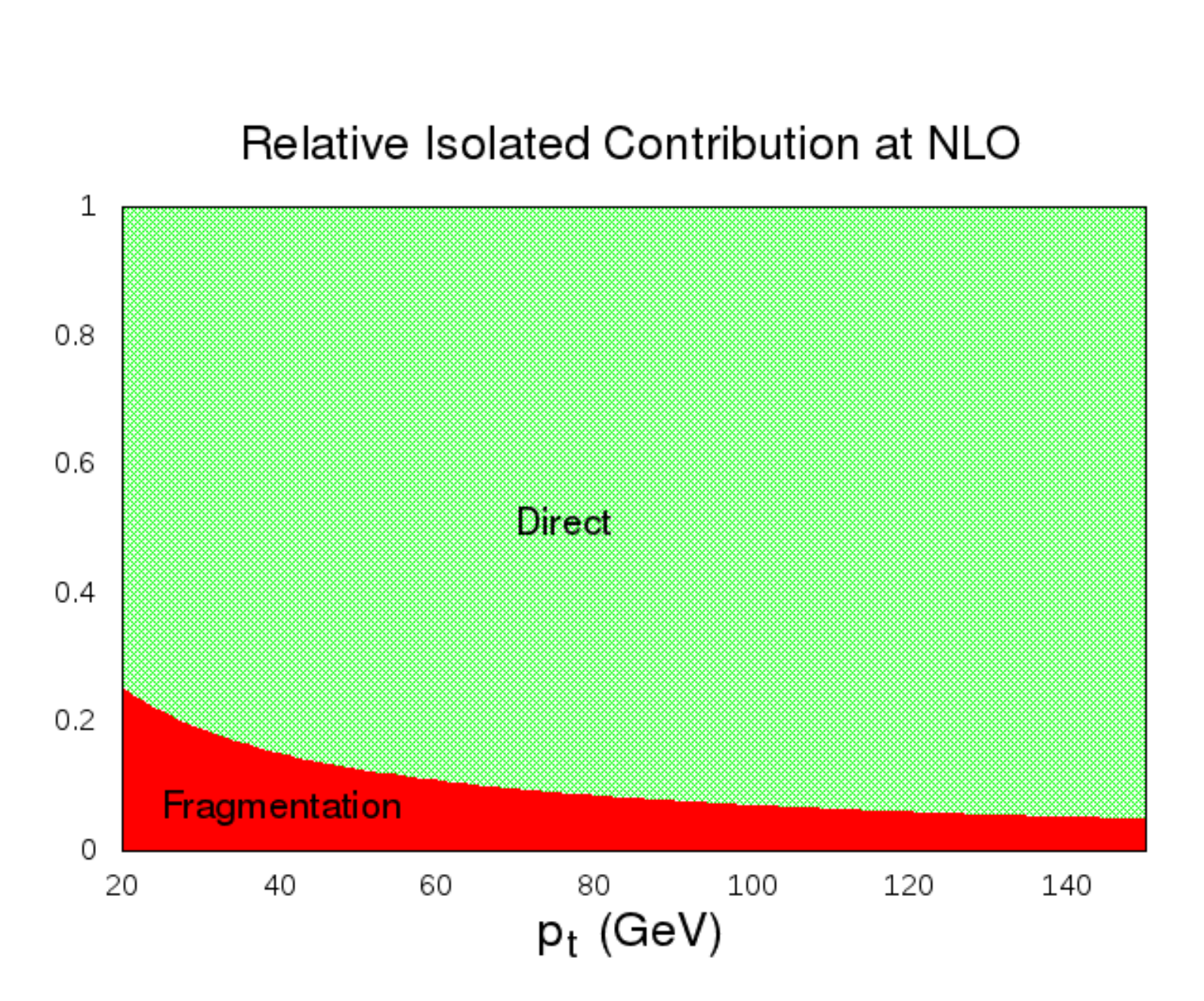}
\caption{These figures show, for DPE prompt photon production at NLO, the relative contributions of the direct and fragmentation processes at a function of photon $p_t$. Left: in the inclusive case, the direct and fragmentation contributions are equal at $p_t\!\simeq\!20 $ GeV, and the relative contribution of direct processes increases with increasing $p_t$. Right: in the isolated case, the direct processes dominate; at $p_t\!\simeq\!20 $ GeV they represent about 75$\%$ of the cross section, and that percentage increases with increasing $p_t$.}
\label{relative_isolated}
\end{figure}

Let us now compare the inclusive and isolated photon production. In Fig.~\ref{relative_isolated} we show in both cases the relative contribution of the direct and the fragmentation processes at NLO. The relative contribution of fragmentation processes is decreasing with increasing $p_t$, but it remains always large in the inclusive case: between 20 and 150 GeV, it goes from 50$\%$ to 25$\%$. In the isolated case however, it is clear that the fragmentation contribution is strongly suppressed by the isolation criteria, and as the transverse momentum of the photon increases, this contribution eventually becomes negligible.

\begin{figure}
\includegraphics[scale=0.4]{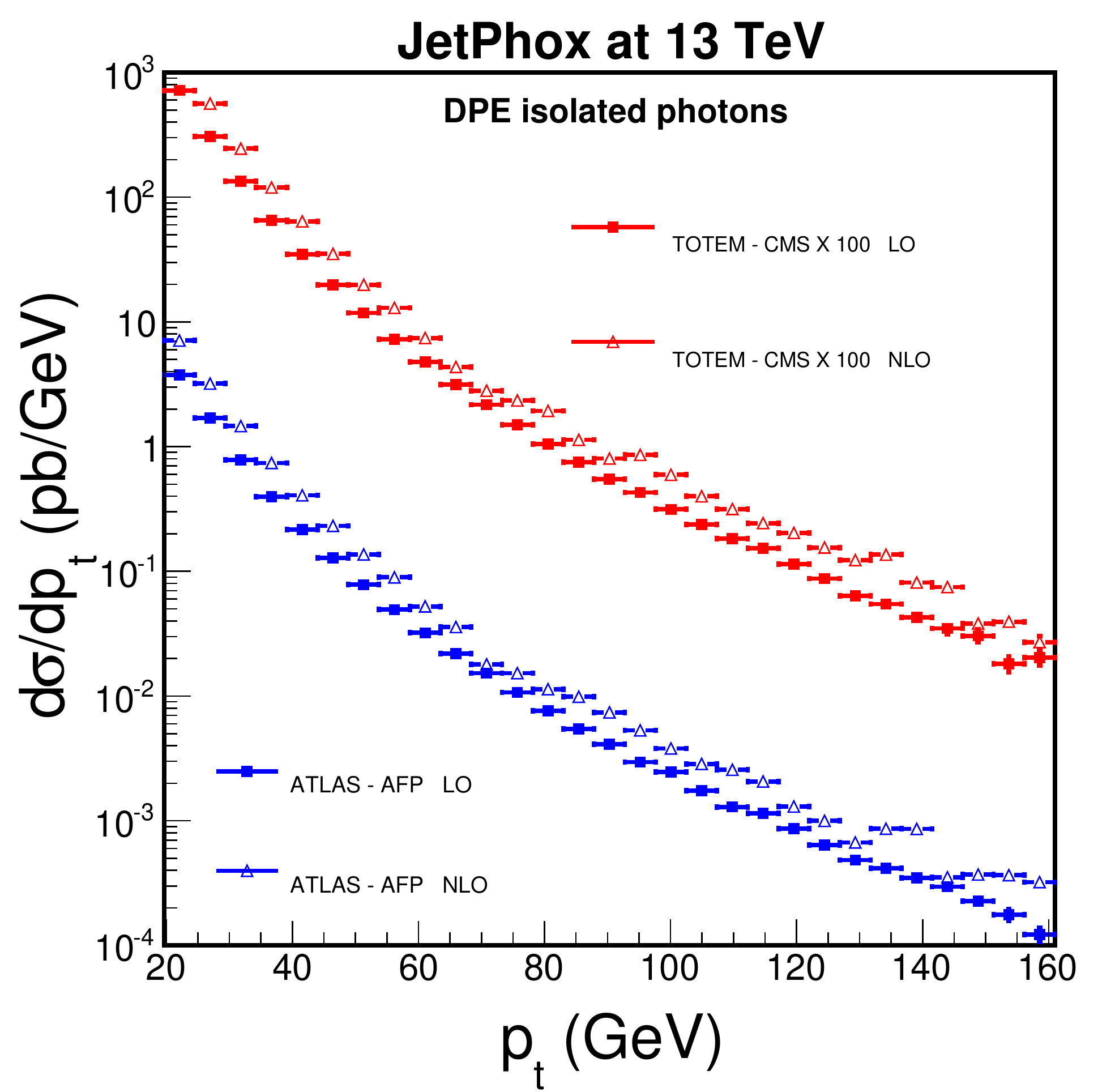}
\includegraphics[scale=0.4]{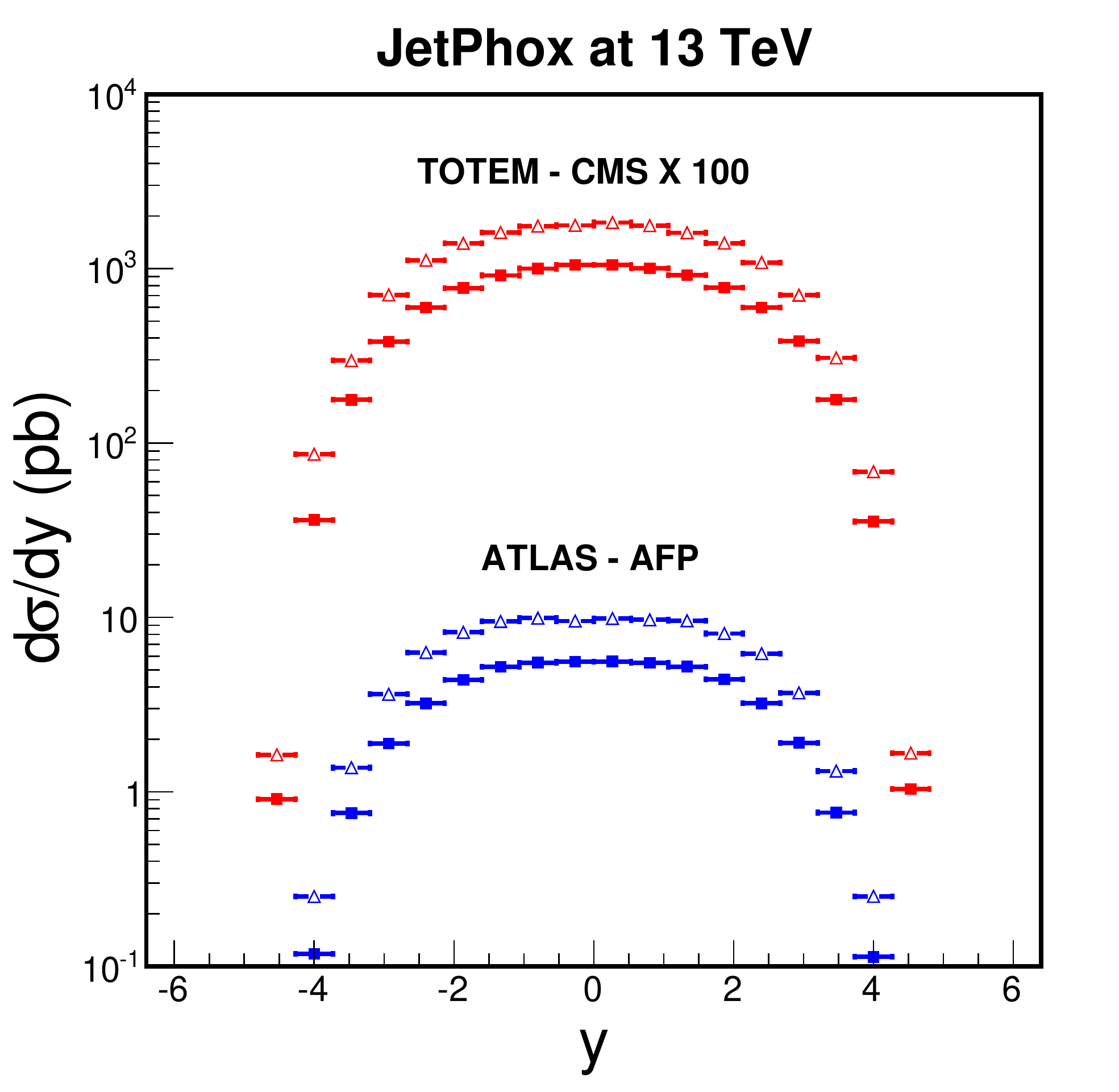}
\caption{These figures show the $p_t$ spectrum (left) and the rapidity distribution (right) of DPE isolated photons for a center of mass energy of 13 TeV, for both ATLAS-AFP and TOTEM-CMS detector acceptances. The differential cross sections at LO (squares) or NLO (triangles) are obtained by summing the direct and fragmentation contributions while requiring that the hadrons measured within a cone of radius $0.4$ around the photon have transverse energy no greater than 4 GeV. The NLO cross sections are about 50$\%$ greater than the LO ones when such isolation criteria is applied.}
\label{distributions_isolated}
\end{figure}

In Fig.~\ref{distributions_isolated}, we display our predictions for the $p_t$ spectrum (left) and the rapidity distribution (right) of DPE isolated photons, for both ATLAS-AFP and TOTEM-CMS detectors at 13 TeV. Comparing the LO and NLO results, we note that the NLO cross sections are about 50$\%$ greater than the LO ones, which is a much bigger difference than in the inclusive case. In order to extract correctly the Pomeron structure from future data, NLO corrections are not even more crucial when the isolation criteria is applied.

\section{Conclusion}

Within the resolved Pomeron model (\ref{colfact}), we have analyzed prompt photon production in DPE processes in p+p collisions. We have performed our calculations of inclusive and isolated photons using the JetPhox program. This is done by substituting the regular pdfs by the diffractive pdfs (\ref{edpdfs}) which also take into account the acceptance of the forward proton detectors. Then, in order to obtain the DPE cross section, we have multiplied the results by the gap survival probability $S_{DPE}$, which we have assumed to be 0.1 for a center of mass energy of 13 TeV.

Using JetPhox allows us to compute the DPE prompt photon production cross sections with, for the first time, next-to-leading order hard matrix elements. Our main result is that the NLO cross sections are larger than the LO ones, by about 20$\%$ in the inclusive case and 50$\%$ in the isolated case. NLO corrections are therefore crucial in such processes at the 13 TeV LHC. In addition, we observed that the isolation criteria is necessary in order to suppress the contribution of fragmentation photons, radiated by high-$p_t$ partons during their fragmentation, and to access in a clean way the direct processes of photon production.

We have also showed that in DPE direct photon production, the Compton partonic sub-processes (Fig~\ref{dpe-photons}-right) clearly dominate over the annihilation ones (Fig~\ref{dpe-photons}-left). This is largely explained by the relative magnitude of the diffractive gluon and quark distribution shown in Fig.~\ref{proton_pdfs}. As a consequence, extracting Pomeron quark distributions from DPE prompt photon measurements will first require that the Pomeron gluon content is already well constrained, which can be done using for instance DPE dijet production \cite{Marquet:2013rja}. 

Finally, we have analyzed different possible scenarios to be tested by LHC experiments ATLAS-AFP and TOTEM-CMS, and we expect that future data on DPE prompt photon production will provide a quantitative way to test the validity of resolved Pomeron model, the factorization of diffractive pdfs into a Pomeron and a Pomeron pdfs, as well as to extract the gap survival probability and understand its behavior with increasing energy. We also hope that $pp\!\to\!pp\gamma X$ measurements at 13 TeV will allow to constraint the quark and gluon structure of the Pomeron, as $pp\!\to\!\gamma X$ measurements have helped constrain regular parton distribution functions \cite{Ichou:2010wc,d'Enterria:2012yj}.

\begin{acknowledgments}

We thank Jean-Philippe Guillet, Christophe Royon and Maria Ubiali for useful discussions and comments. A.K.K. thanks the CAPES-Brazil agency for financial support.

\end{acknowledgments}


\begin{thebibliography}{10}

\bibitem{Derrick:1993xh}
  M.~Derrick {\it et al.} [ZEUS Collaboration],
  Phys.\ Lett.\ B {\bf 315} (1993) 481.

\bibitem{Ahmed:1994nw}
  T.~Ahmed {\it et al.} [H1 Collaboration],
  Nucl.\ Phys.\ B {\bf 429} (1994) 477.

\bibitem{Abachi:1994hb}
  S.~Abachi {\it et al.} [D0 Collaboration],
  Phys.\ Rev.\ Lett.\  {\bf 72} (1994) 2332.
  
\bibitem{Abe:1994de}
  F.~Abe {\it et al.} [CDF Collaboration],
  Phys.\ Rev.\ Lett.\  {\bf 74} (1995) 855.

\bibitem{Chekanov:2008fh} 
  S.~Chekanov {\it et al.}  [ZEUS Collaboration],
  Nucl.\ Phys.\ B {\bf 816}, 1 (2009).

\bibitem{Aaron:2012ad} 
  F.~D.~Aaron {\it et al.}  [H1 Collaboration],
  Eur.\ Phys.\ J.\ C {\bf 72}, 2074 (2012).

\bibitem{Aaron:2012hua}
  F.~D.~Aaron {\it et al.} [H1 and ZEUS Collaborations],
  Eur.\ Phys.\ J.\ C {\bf 72} (2012) 2175.

\bibitem{Collins:1997sr} 
  J.~C.~Collins,
  Phys.\ Rev.\ D {\bf 57}, 3051 (1998)
  [Erratum-ibid.\ D {\bf 61}, 019902 (2000)].

\bibitem{Affolder:2000vb}
  T.~Affolder {\it et al.} [CDF Collaboration],
  Phys.\ Rev.\ Lett.\  {\bf 84} (2000) 5043.

\bibitem{Ingelman:1984ns}
  G.~Ingelman and P.~E.~Schlein,
  Phys.\ Lett.\ B {\bf 152} (1985) 256.

\bibitem{Marquet:2013rja}
  C.~Marquet, C.~Royon, M.~Saimpert and D.~Werder,
  Phys.\ Rev.\ D {\bf 88} (2013) 7,  074029.
  
\bibitem{fpmc} 
  M.~Boonekamp, A.~Dechambre, V.~Juranek, O.~Kepka, M.~Rangel, C.~Royon and R.~Staszewski,
  arXiv:1102.2531 [hep-ph].

\bibitem{Mariotto:2013kca}
  C.~Brenner Mariotto and V.~P.~Goncalves,
  Phys.\ Rev.\ D {\bf 88} (2013) 7,  074023.

\bibitem{jetphox}
  S.~Catani, M.~Fontannaz, J.~P.~Guillet and E.~Pilon,
  JHEP {\bf 0205} (2002) 028;\\
  P.~Aurenche, M.~Fontannaz, J.~P.~Guillet, E.~Pilon and M.~Werlen,
  Phys.\ Rev.\ D {\bf 73} (2006) 094007.

\bibitem{Owens:1986mp}
  J.~F.~Owens,
  Rev.\ Mod.\ Phys.\  {\bf 59} (1987) 465.

\bibitem{afp}
ATLAS Collaboration, Letter of Intent for the Phase-I Upgrade of the ATLAS Experiment, CERN-LHCC-2011-012.

\bibitem{Cisek:2009hp}
  A.~Cisek, W.~Schafer and A.~Szczurek,
  Phys.\ Rev.\ D {\bf 80} (2009) 074013.

\bibitem{Kubasiak:2011xs}
  G.~Kubasiak and A.~Szczurek,
  Phys.\ Rev.\ D {\bf 84} (2011) 014005.

\bibitem{dglap}
  G.~Altarelli and G.~Parisi,
  Nucl.\ Phys.\  B {\bf 126}, 298 (1977);\\
  V.~N.~Gribov and L.~N.~Lipatov,
  Sov.\ J.\ Nucl.\ Phys.\  {\bf 15}, 438 (1972);
  Sov.\ J.\ Nucl.\ Phys.\  {\bf 15}, 675 (1972);\\
  Y.~L.~Dokshitzer,
  Sov.\ Phys.\ JETP {\bf 46}, 641 (1977).

\bibitem{Aktas:2006hy} 
  A.~Aktas {\it et al.}  [H1 Collaboration],
  Eur.\ Phys.\ J.\ C {\bf 48}, 715 (2006).

\bibitem{Trzebinski:2014vha}
  M.~Trzebi\'nski,
  Proc.\ SPIE Int.\ Soc.\ Opt.\ Eng.\  {\bf 9290} (2014) 929026.

\bibitem{lhapdf}
  M.~R.~Whalley, D.~Bourilkov and R.~C.~Group,
  hep-ph/0508110.
  
\bibitem{Khoze:2000cy}
  V.~A.~Khoze, A.~D.~Martin and M.~G.~Ryskin,
  Eur.\ Phys.\ J.\ C {\bf 14} (2000) 525.
  
\bibitem{Kaidalov:2001iz}
  A.~B.~Kaidalov, V.~A.~Khoze, A.~D.~Martin and M.~G.~Ryskin,
  Eur.\ Phys.\ J.\ C {\bf 21} (2001) 521.
  
\bibitem{Bartels:2006ea}
  J.~Bartels, S.~Bondarenko, K.~Kutak and L.~Motyka,
  Phys.\ Rev.\ D {\bf 73} (2006) 093004.
  
\bibitem{Luna:2006qp}
  E.~G.~S.~Luna,
  Phys.\ Lett.\ B {\bf 641} (2006) 171.
  
\bibitem{Frankfurt:2006jp}
  L.~Frankfurt, C.~E.~Hyde, M.~Strikman and C.~Weiss,
  Phys.\ Rev.\ D {\bf 75} (2007) 054009.
  
\bibitem{Gotsman:2007ac}
  E.~Gotsman, E.~Levin and U.~Maor,
  arXiv:0708.1506 [hep-ph].
  
\bibitem{Achilli:2007pn}
  A.~Achilli, R.~Hegde, R.~M.~Godbole, A.~Grau, G.~Pancheri and Y.~Srivastava,
  Phys.\ Lett.\ B {\bf 659} (2008) 137.
  
\bibitem{Khoze:2008cx}
  V.~A.~Khoze, A.~D.~Martin and M.~G.~Ryskin,
  Eur.\ Phys.\ J.\ C {\bf 55} (2008) 363.
  
\bibitem{Ichou:2010wc}
  R.~Ichou and D.~d'Enterria,
  Phys.\ Rev.\ D {\bf 82} (2010) 014015.
  
\bibitem{d'Enterria:2012yj}
  D.~d'Enterria and J.~Rojo,
  Nucl.\ Phys.\ B {\bf 860} (2012) 311.

\end{thebibliography}
\end{document}